\documentclass[11pt,reqno]{amsart}  

\usepackage{amscd} 
\usepackage{mathrsfs}
\usepackage{amsmath,amsbsy,amsthm,amsfonts,amssymb,esint}
\usepackage[unicode,psdextra]{hyperref}
\usepackage{graphicx}
\usepackage{tikz}
\usepackage{bm}
\usepackage[sort&compress,capitalise,nameinlink]{cleveref}
\crefname{section}{\textsection}{\textsection}
\crefname{subsection}{\textsection}{\textsection}
\usepackage{dsfont}
\usepackage{enumerate}

\DeclareMathAlphabet\mathbfcal{OMS}{cmsy}{b}{n}

\newtheorem{theorem}{Theorem}[section]
\newtheorem{lemma}[theorem]{Lemma}
\newtheorem{corollary}[theorem]{Corollary} 
\theoremstyle{definition}

\newtheorem{assumption}{Assumption}
\newtheorem{remark}[theorem]{Remark}

\numberwithin{equation}{section}

\newcommand{\dom}{\mathscr{D}}
\newcommand{\diff}{\mathrm{d}}
\newcommand{\aav}{\mathbf{A}}
\newcommand{\bbv}{\mathbf{B}}
\newcommand{\ssv}{\mathbf{S}}
\newcommand{\supp}{\mathrm{supp}\,}

\newcommand{\Z}{\mathbb{Z}}

\newcommand{\R}{\mathbb{R}}

\newcommand{\bml}[1]{\begin{multline} #1 \end{multline}}
\newcommand{\bmln}[1]{\begin{multline*} #1 \end{multline*}}

\newcommand{\lf}{\left}
\newcommand{\ri}{\right}
\newcommand{\tx}{\textstyle}
\newcommand{\bra}[1]{\lf\langle #1\ri|}
\newcommand{\ket}[1]{\lf|#1 \ri\rangle}
\newcommand{\mean}[3]{\bra{#1}#2\ket{#3}}

\renewcommand{\geq}{\geqslant}

\title[Deficiency indices for singular magnetic Schr\"{o}dinger operators]{Deficiency indices for singular magnetic Schr\"{o}dinger operators}

\author{Michele Correggi}
\address{Dipartimento di Matematica, Politecnico di Milano, P.zza Leonardo da Vinci, 32, 20133, Milano, Italy}
\email{michele.correggi@gmail.com}
\urladdr{https://sites.google.com/view/michele-correggi}

\author{Davide Fermi}
\address{Dipartimento di Matematica, Politecnico di Milano, P.zza Leonardo da Vinci, 32, 20133, Milano, Italy\\
and Istituto Nazionale di Fisica Nucleare, Sezione di Milano, Italy}
\email{davide.fermi@polimi.it}
\urladdr{https://fermidavide.com}

\begin{document}

\begin{abstract} We show that the deficiency indices of magnetic Schr\"odinger operators with several local singularities can be computed in terms of the deficiency indices of operators carrying just one singularity each. We discuss some applications to physically relevant operators.
\end{abstract}


\maketitle

\section{Introduction and Main Result}

Schr\"odinger operators with singular magnetic fields are ubiquitous in physical models describing condensed matter systems and related phenomena. Just to mention a few: the Aharonov-Bohm and Aharonov-Casher effects \cite{AB59,AC79}, the mechanism of Feshbach resonances \cite{CG10} and its applications to cold atom systems, the emergence of anyonic quasi-particles excitation of a 2D electron gas in the fractional quantum Hall regime \cite{Lu23}. 
It is known that  singularities of the magnetic field might be responsible for the possible occurrence of non-trivial deficiency spaces and the related emergence of different self-adjoint realizations of the associated Schr\"odinger operators, see \cite{AT98,CF20,CF23,DF23,F23,PR11} and references therein. In this note we address this precise question for a generic magnetic Schr\"odinger operator with isolated singularities.

We are here mainly inspired by a previous work \cite{BG85}, also related to previous investigations in \cite{CG78,Pe75,Sv81} and \cite[Chpt. 2.5]{Am81} (see also \cite{Be85,Ka85,Kl80,Ne77} for similar studies in the case of Dirac operators), concerning Schr\"odinger operators with electrostatic singular potentials: taking into account a potential with (countably) many local singularities, \emph{e.g.} $V(\bm{x}) \simeq \sum_{j} v_j\,\delta(\bm{x}-\bm{x}_j) + W(\bm{x})$, for some regular $W$, it is proven in \cite[Thm. 2.5]{BG85} that the deficiency indices of the Schr\"odinger operator $-\Delta + V$ equal the sum of the deficiency indices of the operators ``$-\Delta + v_j\,\delta(\bm{x}-\bm{x}_j) $''. In connection with von Neumann or Krein's theory, this result provides some useful information to know a priori the existence and the number of self-adjoint extensions of the original operator, even when the deficiency spaces of the latter are not directly accessible. 
In \cite{IY06} such criterion is exploited to obtain the deficiency indices for the Schr\"odinger Hamiltonian associated to a quantum particle on a two-dimensional torus punctured by a finite number of Aharonov-Bohm fluxes. The result of \cite{BG85} was later reconsidered in \cite{GMNT16}, where a more abstract formulation is provided. We further mention \cite{FMT07} (see also \cite{PR74}), examining a specific family of scalar potentials with several inverse-square local singularities.

Let us now specify the setting that we aim to study in more detail: we consider a magnetic Schr\"odinger operator in $L^2(\R^d)$ ($d \geqslant 1$) of the form
\[
	(-i \nabla + \aav)^2\,,
\]
where 
\[
\aav \simeq \tx\sum_j \aav_j + \aav_0\,, \qquad \aav_0 \in L^{\infty}_{\mathrm{loc}}(\R^d)\,,
\]
and the $\aav_j$'s are singular in non-intersecting regions $\Xi_j$ of co-dimension $\geqslant 1$. The idea is that the $\aav_j$'s are locally singular vector potentials with a regular behaviour at infinity, while $\aav_0$ may diverge there but is locally regular. Our main result is stated in \cref{thm:defind} and proves that the deficiency indices of the full Hamiltonian can be obtained by summing up the deficiency indices of the locally singular operators $(-i \nabla + \aav_j)^2$ and, possibly, of $(-i \nabla + \aav_0)^2$. The proof relies on the use of a local partition of unity to relate the deficiency spaces of the full operator to those of the locally singular ones. This strategy shares some similarities with other works \cite{Br79,Mo79,Cy82,Si78}, on top of \cite{BG85}.


\subsection{Main results}\label{sec:main}

As anticipated, our purpose is to study the deficiency indices of a magnetic Schr\"odinger operator with several isolated singularities and to express them in terms of those of much simpler operators with magnetic fields comprising only one singularity at a time. Heuristically, the idea is that each singularity contributes to the deficiency spaces adding a subspace with a certain dimension -- the deficiency index of the Schr\"odinger operator with that singularity alone --  and therefore the deficiency indices of the whole operator are obtained by summing up the dimension of each local deficiency space. We are going to discuss some examples to which our results apply at the end of this section, but we anticipate that a paradigmatic case satisfying our assumptions is given by singular magnetic Aharonov-Bohm  fluxes sitting at different points in $ \R^2 $.

 We are then interested in characterizing the behaviour of the magnetic Schr\"odinger operator 
	\begin{equation*}
		\dot{H} = (-i \nabla + \aav)^2,		
	\end{equation*}	
where $ \aav $ contains (countably many) local singularities supported in non-intersecting compact sets $ \Xi_j $, $ j \in J $, of zero Lebesgue measure in $ d \geq 1 $ dimensions. Concerning $ \aav $ and such singular sets we make the assumptions below.

	\begin{assumption}
		\label{ass}
		The family of compact sets $ \lf\{ \Xi_j \ri\}_{j \in J} $ and the vector potential $\aav$ are assumed to satisfy the following properties:
	\begin{enumerate}[(H1)]
		\item $J \subseteq \mathbb{N}$ (countably-many singular sets);
	
		\item $\Xi_j \subset \R^d$ is a compact subset of zero Lebesgue measure for any $j \in J$ and we set
			\begin{equation*}
				\Xi := {\tx \bigcup_{j \,\in\, J}}\, \Xi_j\,;
			\end{equation*}

		\item 	there exists $r > 0$ such that
			\begin{equation}\label{eq:disjsupp}
				\mbox{dist}\{\Xi_j,\Xi_{j'}\} \geqslant r\,, \qquad
				\mbox{for all } j \neq j' \in J\,;
			\end{equation}		
		
		\item $ \aav \in L^{\infty}_{\mathrm{loc}}(\R^d \setminus \Xi; \R^d)$ is a real-valued vector potential such that $\nabla \cdot \aav \in L^{\infty}_{\mathrm{loc}}(\R^d \setminus \Xi)$.

	\end{enumerate}
\end{assumption}

	The goal is to investigate the relation between $ \dot{H} $ and the Schr\"odinger operators $ H_j $ containing just one singularity supported on $ \Xi_j $ and which are understandably much easier to deal with. Let us then consider the symmetric realization of $ \dot{H} $ with domain $ \dom(\dot{H}) := C^{\infty}_{\mathrm{c}}(\R^d \!\setminus\! \Xi)\, $ and its closure
	\begin{equation}\label{eq: defHHj}
		H := \overline{\dot{H}}\,.
	\end{equation}
We further refer to the adjoint operator $H^{*}$ with domain
	\begin{equation} \label{eq: defDHst}
		\dom(H^{*}) = \big\{ \psi \!\in\! L^2(\R^d) \;\big|\; (-i \nabla + \aav)^2 \psi \in L^2(\R^d) \big\} \,.
	\end{equation}
In particular, we shall be concerned with the deficiency indices
	\begin{equation*}
		n_{\pm}(H) := \dim\! \big[ \ker(H^{*} \!\mp i) \big]\,.
	\end{equation*}

	\begin{theorem}[Deficiency indices for magnetic Schr\"{o}dinger operators]
		\label{thm:defind}
		\mbox{}\\
		Let \emph{(H1)~--~(H4)} in \cref{ass} hold. Let also $\lf\{ \aav_j \ri\}_{j \in J}, \aav_0 $ be a family of real-valued magnetic potentials such that $ \aav_{j} \in L^{\infty}_{\mathrm{loc}}(\R^d \setminus \Xi_j; \R^d) $, $ \nabla \cdot \aav_{j} \in L^{\infty}_{\mathrm{loc}}(\R^d \setminus \Xi_j) $, for any $ j \in J$,  $ \aav_{0} \in L^{\infty}_{\mathrm{loc}}(\R^d; \R^d) $, $ \nabla \cdot \aav_{0} \in L^{\infty}_{\mathrm{loc}}(\R^d) $ and
			\begin{equation}\label{eq:AminusAj}
				\aav - \aav_0 - \tx\sum_{j \in J} \aav_j \in L^{\infty}(\R^d;\R^d)\,, \qquad
				\nabla \cdot \lf(\aav - \aav_0 - \tx\sum_{j \in J} \nabla \cdot \aav_j \right) \in L^{\infty}(\R^d)\,.
			\end{equation}
		Let $ \dot{H}_j : = ( - i \nabla + \aav_j )^2 $ with domain $ \dom(\dot{H}_j) := C^{\infty}_{\mathrm{c}}(\R^d \!\setminus\! \Xi_j)\, $ and $ H_j $ be its closure \emph{(}with $\Xi_0 := \varnothing$\emph{)}. Let also $ n_{\pm}(H_j) := \dim\!\big[ \ker(H_j^{*} \!\mp i) \big] $ be the associated deficiency indices. Then,
			\begin{equation}\label{eq:npmSum}
				n_{\pm}(H) = n_{\pm}(H_0) + {\tx \sum_{j \in J}}\, n_{\pm}(H_j)\,.
			\end{equation}
	\end{theorem}

	\begin{remark}[Behaviour at infinity]
		\label{rem: infinity}
		\mbox{}	\\
		Since we made no assumption on the behaviour of $\aav$ at infinity, we have to take into account the possible presence of singularities there by extracting the deficiency indices of the operator $H_0 = (-i \nabla + \aav_0)^2$, where $ \aav_0 $ is a regular magnetic potential which possibly has an unbounded support and may diverge at infinity. In typical examples, like a magnetic field which is uniform outside of a compact set, the deficiency indices $n_{\pm}(H_0)$ are actually zero.
	\end{remark}

	\begin{remark}[Real-valuedness of $ \aav $]
		\label{rem: real}
		\mbox{}	\\
		In the above \cref{thm:defind} we assumed that $ \aav $ is real-valued in order to ensure that the original operator $ H $ admits at least one self-adjoint extension -- namely, the Friedrichs' one -- thanks to its positivity (see, \emph{e.g.}, \cite[\S 13.3]{Sc12}).
	\end{remark}
	
	\begin{remark}[Infinite deficiency indices]
		\mbox{}	\\
		The identity \eqref{eq:npmSum} makes evident that $H$ has infinite deficiency indices if and only if at least one of the following two conditions is satisfied:
		\begin{enumerate}[(i)]
			\item there exists at least one $j \in J \cup \{0 \}$ such that $H_j$ has infinite deficiency indices; 
			
			\item $J$ has infinite cardinality and the series on the r.h.s. of \eqref{eq:npmSum} is divergent.
		\end{enumerate}
	\end{remark}

	\cref{thm:defind} applies only to singular magnetic perturbations of the Laplacian, but, combining it with \cite[Thm. 2.5]{BG85}, one can deduce the general result reported hereafter.

	\begin{corollary}[Deficiency indices for generic perturbations]
	\label{thm:genAV}
	\mbox{}\\
	Let the assumptions of \cref{thm:defind} hold and let $V \in L^2_{\mathrm{loc}}(\R^d \setminus \Xi)$ be a real-valued electrostatic potential bounded from below. Let also $ \lf\{ \aav_j \ri\}_{j \in J}, \aav_0 $ be a family of magnetic potentials as in \cref{thm:defind} and $\{V_j\}_{j \in J}, V_0$ be a family of electrostatic potentials bounded from below such that $V_j \in L^2_{\mathrm{loc}}(\R^d \setminus \Xi_j)$, for any $j \in J$, $V_0 \in L^2_{\mathrm{loc}}(\R^d)$ and
	\begin{equation*}
		V - V_0 - \tx \sum_{j \in J} V_j \in L^2(\R^d)\,.
	\end{equation*}
	Consider the Schr\"odinger operators 
		\begin{equation*}
			H_V := (-i \nabla + \aav)^2 + V  \,, \qquad 			H_{V,j} := (-i \nabla + \aav_j)^2 + V_j \,,
		\end{equation*}
		given by the closure of the symmetric realizations on $ C^{\infty}_{\mathrm{c}}(\R^d \setminus \Xi) $ and $ C^{\infty}_{\mathrm{c}}(\R^d \setminus \Xi_j) $ \emph{(}with $\Xi_0 := \varnothing$\emph{)}, respectively. Let also $ n_{\pm}(H_V) $ and $ n_{\pm}(H_{V,j}) $ be the associated deficiency indices.	 Then,
		\begin{equation*}
			n_{\pm}(H_V) = n_{\pm}(H_{V,0}) + \tx\sum_{j \in J} n_{\pm}(H_{V,j})\,.
		\end{equation*}
	\end{corollary}
	
	\begin{remark}[Real-valuedness and lower-boundedness of $ V $]
		\mbox{}	\\
		As in \cref{rem: real}, the hypothesis that $ V $ is real and bounded from below is made to guarantee that $ H_V $ has self-adjoint extensions, due to the fact that it is bounded from below.
	\end{remark}

\subsection{Physical examples}
We now discuss some physical models satisfying our assumptions and the related consequences of our main result. Consider then a charged spinless quantum particle moving in presence of many parallel straight cylindrical solenoids. In particular, let us focus on a low-energy regime where the wave-length of the particle is much larger than the solenoids diameter and, at the same time, much smaller than their length. With this physical setting in mind, we henceforth refer to the idealized configuration involving solenoids of zero diameter and infinite length. Accordingly, by exploiting the invariance under translations along the axial direction and an obvious factorization, we can reduce the analysis to a 2D problem.

More precisely, let $J \subseteq \mathbb{N}$ be the (finite or at most countably infinite) set of indices labeling the solenoids and let $\{ \mathbf{x}_j\}_{j \in J} \subset \R^2$ be the set of points identifying their positions. We assume that
	\begin{equation}\label{eq:defr}
		\inf_{j \neq j' \in J} |\mathbf{x}_j - \mathbf{x}_{j'}| \geqslant r, \qquad \mbox{for some fixed\, $r > 0$}\,.
	\end{equation}
The dynamics of the particle is described by the Schr\"odinger operator in $\R^2$ given by
	\begin{equation}\label{eq:HmanyAB}
		H = (- i\nabla + \aav)^{2} , \qquad \aav(\mathbf{x}) = {\tx \sum_{j \in J}}\, \alpha_j\,\tfrac{(\mathbf{x} - \mathbf{x}_j)^{\perp}}{|\mathbf{x} - \mathbf{x}_j|^2}\,,
	\end{equation}
where, for each $j \in J$, $\alpha_j \in \R \setminus \mathbb{Z}$ is a parameter related to the magnetic flux across the $j$-th solenoid and we have introduced the notation $ \mathbf{x}^{\perp} = (-x_2,x_1)$ for any $ \mathbf{x} = (x_1,x_2) \in \R^2$. We understand $H$ to be initially defined as a symmetric operator on the graph closure of $C^{\infty}_{\mathrm{c}}\big(\R^2 \setminus \{ \mathbf{x}_j\}_{j \in J}\big)$.

To make connection with the analysis outlined in  \cref{sec:main}, it is natural to fix $\Xi_j = \{ \mathbf{x}_{j}\}$ for $j \in J$ and to observe that all the conditions in \cref{ass} are indeed satisfied. Furthermore, a family of magnetic potentials as in the statement of \cref{thm:defind} is given by
	\begin{equation} \label{eq: aavj}
		\aav_j = \alpha_j \tfrac{(\mathbf{x} - \mathbf{x}_j)^{\perp}}{\lf| \mathbf{x} - \mathbf{x}_j \ri|}\,,	\qquad \mbox{for } j \in J\,.
	\end{equation}
We denote with $ H_j $ the associated single-flux Aharonov-Bohm operators (which are the closures of the symmetric realizations of)
	\begin{equation*}
		{H}_j = \lf(- i\nabla + \aav_j \ri)^2.
	\end{equation*}
Recalling that $\alpha_j \in \R \setminus \mathbb{Z}$ for all $j \in J$, each of the latter operators has finite deficiency indices  \cite{AT98,DS98}
	\begin{equation*}
		n_{\pm}(H_j) = 2\,, \qquad \mbox{for all\, $j \in J$}\,.
	\end{equation*}
Keep also in mind that the free Laplacian $H_0 = - \Delta$ on $C^{\infty}_{\mathrm{c}}(\R^2)$ has zero deficiency indices.
	
Taking the above arguments into account, from \cref{thm:defind} we deduce
	\begin{equation}\label{eq:npmAB}
		n_{\pm}(H) = {\tx \sum_{j \in J}}\, n_{\pm}(H_j) = 2 |J|\,,
	\end{equation}
understanding that $H$ has infinite deficiency indices if $J$ has countably infinite cardinality. This means that $H$ admits exactly $2|J|$ self-adjoint extensions, by standard von Neumann theory. In the case where $J$ is finite, the same conclusion was derived in \cite{CF23} by means of resolvent techniques.

	\begin{remark}[Magnetic traps]
		\mbox{}	\\
		Assuming that the number of singularities is finite ($|J| < \infty$), the above result could be generalized to the case where, on top of the Aharonov-Bohm fluxes, a real-valued regular potential $\ssv \in L_{\mathrm{loc}}^{\infty}(\R^2;\R^2) $ is also present (compare with the results in \cite{CF20,F23}), possibly describing a magnetic trapping. In this configuration, the reference operator becomes $H_0 = (-i \nabla + \ssv)^2$ on $C^{\infty}_{\mathrm{c}}(\R^2)$, instead of the free Laplacian. Notice that the deficiency indices $n_{\pm}(H_0)$ may not vanish, depending on the behaviour of $ \ssv $ at infinity. Considering this, \eqref{eq:npmAB} generalizes to
		\begin{equation*}
			n_{\pm}(H) = n_{\pm}(H_0) + 2 |J|\,.
		\end{equation*}
		Requiring $J$ to have finite cardinality is here crucial in order to avoid the simultaneous occurrence of singularities $\Xi_j$ accumulating at infinity with a divergent magnetic perturbation $\ssv$. For models of this kind it would be impossible in general to apply our main result, or any straightforward variation thereof, given that there would be no decoupling of the deficiency indices of $H_{0}$ and $H_{j}$ for all $j \in J$.	
	\end{remark}
	
	\begin{remark}[Point interactions]
		\mbox{}	\\
		Assume that some of the solenoids have integer magnetic fluxes $\alpha_j \in \mathbb{Z}$. Let $Y \subset J$ be the set of labels associated to such solenoids. As before we have $n_{\pm}(H_j) = 2$ for any $j \in J \setminus Y$. On the opposite, for any $j \in Y$ the symmetric operator $H_j$ is unitarily equivalent to the Laplacian defined on $C^{\infty}_{\mathrm{c}}(\R^2 \setminus \{ \mathbf{x}_j\})$ and it therefore admits as self-adjoint extensions any one of the usual Hamiltonians with a delta-type interaction concentrated at a point. Accordingly, we have $n_{\pm}(H_j) = 1$ for any $j \in Y$. In this case, \cref{thm:defind} gives
		\begin{equation*}
			n_{\pm}(H) = {\tx \sum_{j \in J \setminus Y}}\, n_{\pm}(H_j) + {\tx \sum_{j \in Y}}\, n_{\pm}(H_j) = 2 \,|J \setminus Y| + |Y|\,.
		\end{equation*}
	\end{remark}
		
	\begin{remark}[Singular electrostatic interactions]
		\mbox{}	\\
		A further generalization concerns a model involving repulsive electrostatic interactions, with singularities centered on top of the Aharonov-Bohm flux points. More precisely, let us fix $\alpha_j \in \R \setminus \mathbb{Z}$, $q_j \in \R $, $p_j \geqslant 0$ for $j \in J$ and consider the Hamiltonian operator
		\begin{equation*}
			H_V = (-i \nabla + \aav)^2 + V\,, \qquad \aav(\mathbf{x}) = {\tx \sum_{j \in J}}\, \alpha_j\,\tfrac{(\mathbf{x} - \mathbf{x}_j)^{\perp}}{|\mathbf{x} - \mathbf{x}_j|^2}\,, \qquad V(\mathbf{x}) = {\tx \sum_{j \in J}}\! \lf( \tfrac{q_j}{|\mathbf{x} - \mathbf{x}_j|} + \tfrac{p_j}{|\mathbf{x} - \mathbf{x}_j|^2}\ri) .
		\end{equation*}
		We are then let to consider the family of operators $H_{V,j} = \lf(- i\nabla + \aav_j\ri)^{\!2} + V_j$, $j \in J$, where $ \aav_j $ is defined in \eqref{eq: aavj} and
		\begin{equation*}
			V_j(\mathbf{x}) =  \frac{q_j}{|\mathbf{x} - \mathbf{x}_j|} + \frac{p_j}{|\mathbf{x} - \mathbf{x}_j|^2} \,.
		\end{equation*}
		By decomposition in angular harmonics one infers that the radial operator related to $ {H}_j$ and associated to the $\ell_j$-th harmonic ($\ell_ j \in \mathbb{Z}$) is unitarily equivalent to
		\begin{equation*}
			 {h}_j := -\frac{d^2}{dr^2} + \frac{(\ell_j + \alpha_j)^2 + p_j - 1/4}{r^2} + \frac{q_j}{r}\qquad \mbox{on\, $C^{\infty}_c(\R_{+})$}\,.
		\end{equation*}
		Building on classical results for such kinds of operators \cite{BDG11,BG85,DFNR20,DR17,DF23}, we deduce that
		\begin{equation*}
			n_{\pm}({h}_j) = \lf\{\!\!\begin{array}{ll}
				1	&	\mbox{if } (\ell_j + \alpha_j)^2 + p_j \in [0,1) \,, \\
				0	&	\mbox{otherwise}\,.
			\end{array}
			\ri.
		\end{equation*}
		It is remarkable that the deficiency indices do not depend on $q_j$. Let us indicate with $\lfloor \alpha_j \rfloor$ the integer part of $\alpha_j$. Then, setting
		\begin{align*}
			J_2 & := \lf\{j \!\in\! J\,\big|\, \max\!\lf\{(\alpha_j \!-\! \lfloor \alpha_j \rfloor)^2,(\alpha_j \!-\! \lfloor \alpha_j \rfloor - 1)^2\ri\} + p_j < 1 \ri\} , \\
			J_1 & := \lf\{j \!\in\! J\,\big|\, \min\!\lf\{(\alpha_j \!-\! \lfloor \alpha_j \rfloor)^2,(\alpha_j \!-\! \lfloor \alpha_j \rfloor - 1)^2\ri\} + p_j < 1,  \max\!\lf\{(\alpha_j \!-\! \lfloor \alpha_j \rfloor)^2,(\alpha_j \!-\! \lfloor \alpha_j \rfloor - 1)^2\ri\} + p_j \geqslant 1 \ri\} , \\
			Y & := J \setminus (J_2 \cup J_1)\,,
		\end{align*}
		we obtain
		\begin{equation*}
			n_{\pm}(H_{V,j}) = \lf\{\!\!\begin{array}{ll}
				2	&	\mbox{if $j \in J_2$}\,, \vspace{0.1cm} \\
				1	&	\mbox{if $j \in J_1$}\,, \vspace{0.1cm} \\
				0	&	\mbox{if $j \in Y$}\,.
			\end{array}\ri.
		\end{equation*}
		Then, by \cref{thm:genAV}, we deduce that
		\begin{equation*}
			n_{\pm}(H_V) = 2\,|J_2| + |J_1|\,.
		\end{equation*}
		In particular, for purely Coulomb interactions with non-integer magnetic fluxes ($p_j = 0$, $\alpha_j \in \R \setminus \Z$, for all $j \in J$) one readily checks that $J = J_2$ ($J_1 = Y = \varnothing$), so that the above relation reduces to
		\begin{equation*}
			n_{\pm}(H) = 2|J| \,.
		\end{equation*}

	\end{remark}

\bigskip

\noindent
{\footnotesize
\textbf{Acknowledgments}
This work has been supported by MUR grant Dipartimento di Eccellenza 2023--2027 of Dipartimento di Matematica, Politecnico di Milano. MC acknowledges the support of PNRR Italia Domani and Next Generation Eu through the ICSC National Research Centre for High Performance Computing, Big Data and Quantum Computing. DF acknowledges the support INdAM-GNFM Progetto Giovani 2020 ``\textsl{Emergent Features in Quantum Bosonic Theories and Semiclassical Analysis}''.}

\section{Proofs}

\subsection{Basic assumptions and definitions}

For notational convenience, in the sequel we set $J_0 = J \cup \{0\}$, understanding $\Xi_0 = \varnothing$.
Let us then refer to the family of vector potentials $\{\aav_{j}\}_{j \in J_0}$ introduced in the statement of \cref{thm:defind}. We remark that the hypothesis (H3) in \cref{ass} and the condition \eqref{eq:AminusAj} in \cref{thm:defind} ensure that the sum $\sum_{j \in J_0} \aav_j$ is well-defined as an element of $L^2_{\mathrm{loc}}(\R^d \setminus \Xi;\R^d)$, also when $J$ is a countable infinite set.
On the other hand, hypotheses (H2) and (H3) ensure that $\Xi$ is closed and has zero Lebesgue measure. The same assumptions further yield the following result, which we report from \cite[Lemma 2.1]{BG85} and relies on the existence of partitions of unity \cite[Corollary 1.4.11]{Ho83}.

	\begin{lemma}\label{lemma:etaj}
		Assume hypotheses \emph{(H1)~--~(H3)}. Then, there exist three families $\{\rho_j\}_{j \in J_0}, \{\eta_j\}_{j \in J_0},\{\tilde{\eta}_j\}_{j \in J_0} \linebreak \subset C^{\infty}(\R^d)$ of cut-off functions fulfilling the following conditions:
		\begin{enumerate}[i)]
	
			\item $\partial^{\alpha} \rho_{j}, \partial^{\alpha} \eta_{j}, \partial^{\alpha} \tilde{\eta}_{j} \in L^{\infty}(\R^d)$ for all multi-indices $\alpha$ with $0 \leqslant |\alpha| \leqslant 2$ and for all $j \in J_0$;
	
			\item the supports of functions in the same family are pairwise disjoints;
	
			\item $\Xi_j \subset \supp(\rho_j) \subset \supp(\eta_j) \subset \supp(\tilde{\eta}_j)$ and 
				\[
					\left.\rho_j \right|_{\Xi_j} = 1\,, \qquad \left.\eta_{j}\right|_{\supp(\rho_j)} = 1\,, \qquad \left. \tilde{\eta}_{j} \right|_{\supp(\eta_j)} = 1\,, \qquad \mbox{for all } j \in J_0 \; ;
				\]
	
			\item there exists $0<\delta < r/2$ such that $\mbox{\emph{dist}}\{\supp (1-\eta_j), \supp(\rho_j)\} \geqslant \delta$ for all $j \in J_0$.

		\end{enumerate}
	\end{lemma}

We use the first partition of unity to define out of the $\aav_j$'s a new family of vector potentials with pairwise disjoint supports and set
	\begin{equation}\label{eq:bbvj}
		\bbv_j := \rho_j\,\aav_j\,, \qquad \mbox{for all $j \in J_0$} \,.
	\end{equation}
Furthermore we introduce a new family of operators:
	\[
		K = \lf(-i \nabla + \tx \sum_{j \in J_0} \bbv_j\ri)^2 , \qquad\quad 
		K_j := (-i \nabla + \bbv_j)^2\,, \quad \mbox{for all $j \in J_0$}\,,
	\]
regarded as the closures of the corresponding formal expressions restricted to smooth functions with supports away from $\Xi$ and $\Xi_j$, respectively. The idea behind the introduction of the two other partitions $ \{\eta_j\}_{j \in J_0}$, $\{\tilde{\eta}_j\}_{j \in J_0} $ is to have partition functions equal to 1 in the support of the magnetic potentials $ \bbv_j $ and decaying to zero in a slightly larger region around $ \Xi_j $ (see \cref{lemma:BG22,lemma:BG23,lemma:BG24}, as well as \cref{rem:tildeeta}).

We notice that, since the vector potentials $(\bbv_j)_{j \in J_0}$ are real-valued, the operators $K$ and $K_{j}$, $j \in J_0$, are certainly non-negative.
As a consequence, they all admit self-adjoint extensions in $L^2(\R^d)$ (see once more \cite[\S 13.3]{Sc12}) and, accordingly, 
			\begin{equation}\label{eq:npnm}
				n_{+}(K) = n_{-}(K)\,, \qquad n_{+}(K_j) = n_{-}(K_j)\,.
			\end{equation}

	\begin{lemma}\label{lemma:SKatosmall}
		$n_{\pm}(H) = n_{\pm}(K)$ and $n_{\pm}(H_j) = n_{\pm}(K_j)$, for all $j \in J_0$.
	\end{lemma}
	
\begin{proof}
We show that, for any choice of cut-off functions $\{\rho_j\}_{j \in J_0}$ as in \cref{lemma:etaj}, the closed symmetric operator $K$ is an infinitesimally Kato-small perturbation of $H$. Similar arguments can be employed to prove that $K_j$ is an infinitesimal perturbation of $H_j$, for all $j \in J_0$. The thesis ultimately follows by a variant of the Kato–Rellich theorem, see \emph{e.g.} \cite[Cor. 2]{BF77}, \cite[Thm. 9, p. 100]{BS87}, and \cite[Eq. (1.1) and related references]{Ki03}.

To prove our claim, let us first point out that, upon setting
	\begin{equation*}
		\ssv := \aav -  \tx\sum_{j \in J_0} \bbv_j \,,
	\end{equation*}
a straightforward computation gives
	\begin{equation*}
		H  = K + 2 \,\ssv \cdot \left(-i \nabla + \tx\sum_{j \in J_0} \bbv_j\right) - i (\nabla \!\cdot\! \ssv) + \ssv^2.
	\end{equation*}
	
Notice that for any $\psi \in C^{\infty}_c(\R^d \setminus \Xi)$ and for all $\varepsilon > 0$, using the Cauchy–Schwarz inequality together with the basic relation $\sqrt{u\,v} \leqslant \varepsilon\,u + \frac{1}{2\varepsilon}\,v$, we get
	\begin{equation*}
		\left\|\left(-i \nabla + \tx\sum_{j \in J_0} \bbv_j\right)  \psi \right\|_{2} = \sqrt{ \mean{\psi }{K}{\psi} } \leqslant \varepsilon\,\|K \psi\|_{2} + \tfrac{1}{2\varepsilon}\,\|\psi\|_{2}\,.
	\end{equation*}
Let us further remark that the hypotheses in \eqref{eq:AminusAj} and the features of $\rho_j$ stated in \cref{lemma:etaj} ensure that $\ssv$ and $\nabla \cdot \ssv$ are indeed uniformly bounded everywhere.
Then, by standard density considerations we obtain, for any $\psi \in \dom(K)$,
	\begin{equation*}
		\lf\|(H - K)\psi\ri\|_{2} \leqslant  2\varepsilon \lf\|\ssv\ri\|_{\infty}\, \lf\|K \psi\ri\|_{2} + \lf(\tfrac{1}{\varepsilon} \lf\|\ssv\ri\|_{\infty} + \lf\|\nabla \!\cdot\! \ssv\ri\|_{\infty} + \lf\|\ssv\ri\|_{\infty}^2 \ri) \lf\|\psi\ri\|_{2}\,,
	\end{equation*}
which, in view of the arbitrariness of $\varepsilon > 0$, ultimately yields the thesis.
\end{proof}

On account of \cref{lemma:SKatosmall}, in the following we study $K$ in place of $H$. Accordingly, we often refer to the adjoint operators $K^{*}$ and $K_j^{*}$, $j \in J_0$, with domains respectively given by
	\begin{align}
		\dom(K^{*}) & = \big\{ \psi \!\in\! L^2(\R^d) \;\big|\; \big(-i \nabla + \tx \sum_{j \in J_0} \bbv_j\big)^2 \psi \in L^2(\R^d) \big\} \,, \label{eq: defDKst} \\
		\dom(K_j^{*}) & = \big\{ \psi \!\in\! L^2(\R^d) \;\big|\; \lf(-i \nabla + \bbv_j\ri)^2 \psi \in L^2(\R^d) \big\} \,. \label{eq: defDKjst}
	\end{align}

	\begin{lemma}\label{lemma:BG22}
		Assume hypotheses \emph{(H1)~--~(H4)} and let $\{\eta_j\}_{j \in J_0}$ be a family of cut-off functions as in \cref{lemma:etaj}. Then, for all $j \in J_0$, the following implications hold true:
		\begin{gather}
			\psi \in \dom(K_{j}^{*})	\quad \Longrightarrow \quad \eta_j \psi \in \dom(K_{j}^{*}) \cap \dom(K^{*})\,; \label{eq:etajpsiDHjst}\\
			\psi \in \dom(K^{*})	\quad \Longrightarrow \quad \eta_j \psi \in \dom(K_{j}^{*}) \cap \dom(K^{*})\,.
		\end{gather}
		Moreover, for $\psi \in \dom(K_{j}^{*})$ or $\psi \in \dom(K^{*})$,
		\begin{equation}
			K^{*}(\eta_j \psi) = K_{j}^{*}(\eta_j \psi) 
			= \eta_j (- i \nabla \!+\! \bbv_j)^2 \psi - 2(\nabla \eta_j) \!\cdot\! (\nabla \psi) + (-\Delta \eta_j) \psi \,. \label{eq:Hjetajpsi}
		\end{equation}
	\end{lemma}
	
\begin{proof}
As a foreplay, we show that $\dom(K_{j}^{*}) \subset H^2_{\mathrm{loc}}(\R^d \setminus \supp \bbv_j)$ for any $j \in J_0$. To this avail, let $\psi \in \dom(K_{j}^{*})$ and notice that the very definition \eqref{eq: defDKjst} of $\dom(K_{j}^{*})$ yields $\xi_j (-i \nabla + \bbv_j)^2 \psi = \xi_j (-\Delta \psi) \in L^2(\R^2)$ for any $\xi_j \in C^{\infty}_{\mathrm{c}}(\R^d \setminus \supp \bbv_j)$, whence $\Delta \psi \in L^2_{\mathrm{loc}}(\R^d \setminus \supp \bbv_j)$. From here we deduce $\psi \in H^2_{\mathrm{loc}}(\R^d \setminus \supp \bbv_j)$ by standard elliptic regularity, see \emph{e.g.} \cite[p.~125, Thm. 3.2]{LM72}.
It can be proved in a similar manner that $\dom(K^{*}) \subset H^2_{\mathrm{loc}}(\R^d \setminus \bigcup_{j \in J_0}\, \supp \bbv_j)$.

Now, let us fix $j \in J_0$ and take $\psi \in \dom(K_{j}^{*})$. Let also $\eta_j \in C^{\infty}(\R^d)$ be the associated cut-off function as in \cref{lemma:etaj} and notice that $\eta_j \big|_{\supp \bbv_j} = 1$, since $\supp \bbv_j \subseteq \supp \rho_j$ (see \eqref{eq:bbvj}). Then, $\eta_j \psi \in L^2(\R^d)$ and, by direct inspection, we infer
	\bml{\label{eq:Hetajpsi}
		\lf(- i \nabla + \tx \sum_{k \in J_0} \bbv_k\ri)^2 (\eta_j \psi) = (- i \nabla + \bbv_j)^2 (\eta_j \psi) \\
		= (-\Delta \eta_j) \psi + 2\, (-i \nabla \eta_j) \cdot (- i \nabla \psi) + \eta_j (- i \nabla + \bbv_j)^2 \psi \,\in\, L^2(\R^d)\,,
	}
where we have taken into account that $\eta_j,\Delta \eta_j \in L^\infty(\R^d)$, $\nabla \eta_j \in L^\infty(\R^d;\R^d)$ with $\supp(\nabla \eta_j) \subset \R^d \setminus \supp \bbv_j$, together with $\psi \in H^2_{\mathrm{loc}}(\R^d \setminus \supp \bbv_j)$ and $(- i \nabla + \bbv_j)^2 \psi \in L^2(\R^d)$.

By the same token it can be inferred that, for any $\psi \in \dom(K^{*})$ and $\eta_j$ as before, there holds
	\bml{\label{eq:Hetapsi}
		(- i \nabla + \bbv_j)^2 (\eta_j \psi) = \lf(- i \nabla + \tx \sum_{k \in J_0}\bbv_k \ri)^2 (\eta_j \psi) \\
		= (-\Delta \eta_j) \psi + 2 (-i \nabla \eta_j) \cdot \lf(- i \nabla + \tx \sum_{k \in J_0}\bbv_k\ri) \psi + \eta_j \lf(- i \nabla + \tx \sum_{k \in J_0}\bbv_k \ri)^2 \psi \,\in\, L^2(\R^d)\,.
	}
The thesis ultimately follows, recalling once more the definitions \eqref{eq: defDKst} and \eqref{eq: defDKjst} of $\dom(K^{*})$ and $\dom(K_{j}^{*})$.
\end{proof}

	\begin{lemma}\label{lemma:BG23}
		Assume hypotheses \emph{(H1)~--~(H4)} and let $\{\eta_j\}_{j \in J_0}$ be a family of cut-off functions as in \cref{lemma:etaj}. Then, for all $j \in J_0$, the following implications hold true:
		\begin{gather}
			\psi \in \dom(K_j) \quad\; \Longrightarrow \quad\; \eta_j \psi \in \dom(K_{j}) \cap \dom(K)\,; \label{eq:etajpsiDHj}\\
			\psi \in \dom(K)	\quad\; \Longrightarrow \quad\; \eta_j \psi \in \dom(K_{j}) \cap \dom(K)\,. \label{eq:etajpsiDH}
		\end{gather}
		Moreover, for $\psi \in \dom(K_j)$ or $\psi \in \dom(K)$,
		\begin{equation}
			K(\eta_j \psi) = K_{j}(\eta_j \psi) 
			= \eta_j (- i \nabla \!+\! \bbv_j)^2 \psi - 2(\nabla \eta_j) \!\cdot\! (\nabla \psi) + (-\Delta \eta_j) \psi \,.
		\end{equation}
	\end{lemma}
	
	\begin{remark}[Partition  $\{\tilde{\eta}_j\}_{j \in J_0} $]
		\label{rem:tildeeta}
		\mbox{}	\\
		Note that under the conditions of \cref{lemma:etaj}, the result of \cref{lemma:BG23} immediately applies also to the family of cut-off functions $  \{\tilde{\eta}_j\}_{j \in J_0} $. We are going to exploit this fact in the proof of \cref{thm:defind}.
	\end{remark}
	
\begin{proof}

We discuss the proof of \eqref{eq:etajpsiDHj} as an example. Similar arguments can be employed to derive \eqref{eq:etajpsiDH}.
Let us firstly recall that the domain $\dom(K_j)$ is by definition the closure in the graph-norm topology of the dense set $C^{\infty}_{\mathrm{c}}(\R^d \setminus \Xi_j)$. So,
	\begin{equation*}
		\forall\,\psi \!\in\! \dom(K_j), \;\; \exists\, \{\psi_n\}_{n \in \mathbb{N}} \subset C^{\infty}_{\mathrm{c}}(\R^d \setminus \Xi_j) \quad \mbox{ s.t. } \quad \psi_n \! \xrightarrow[n \to + \infty]{L^2(\R^d)} \psi, \quad K_j \psi_n \xrightarrow[n \to + \infty]{L^2(\R^d)} K_j \psi \, .
	\end{equation*}
Since $\eta_j \in L^{\infty}(\R^d)$, it follows immediately that $\eta_j \psi_n \to \eta_j \psi$ in $L^2(\R^d)$. On the other side, we have
	\begin{equation}\label{eq: Hetajpsin}
		K_j(\eta_j \psi_n) 
		= (- \Delta \eta_j) \psi_n + 2 (-i \nabla \eta_j) \cdot (-i \nabla + \bbv_j) \psi_n +\eta_j K_j \psi_n \,.
	\end{equation}
Notice that $(- \Delta \eta_j) \psi_n \to (- \Delta \eta_j) \psi$ and $\eta_j K_j \psi_n \to \eta_j K_j \psi$ in $L^2(\R^d)$. Let us now examine the second term on the r.h.s. of \eqref{eq: Hetajpsin}. We remark that $\psi \in \dom(K_j) \subset \dom(K_j^*) \subset H^2_{\mathrm{loc}}(\R^d \setminus \supp \bbv_j)$ (see the proof of \cref{lemma:BG22}) and $\nabla \eta_j \in C^{\infty}_{\mathrm{c}}(\R^d \setminus \supp \bbv_j;\R^d)$. Taking this into account, we infer $(-i \nabla \eta_j) \cdot (-i \nabla + \bbv_j) \psi = (\nabla \eta_j) \cdot (\nabla \psi) \in L^2(\R^d)$. Moreover, for any given $\xi \in C^{\infty}_{\mathrm{c}}(\R^d \setminus \supp \bbv_j)$, an integration by parts and elementary estimates yield
	\bmln{
		\lf\| \xi (-i \nabla + \bbv_j) (\psi_n - \psi) \ri\|_{2}^{2} \\
		= \int_{\R^d}\!\! \diff\mathbf{x}\; (\psi_n \!- \psi)^{*}\, (-i \nabla \xi^2) \cdot (-i \nabla + \bbv_j) (\psi_n\! - \psi)
			+ \int_{\R^d}\!\! \diff\mathbf{x}\; \xi^2\, (\psi_n\! - \psi)^{*}\, (-i \nabla + \bbv_j)^2 (\psi_n \!- \psi) \\
		\leqslant 2\, \lf\|\nabla \xi \ri\|_{\infty}\, \lf\|\psi_n - \psi \ri\|_{2}\, \lf\| \xi (-i \nabla + \bbv_j) (\psi_n - \psi)\ri\|_{2}
			+ \lf\|\xi \ri\|_{\infty}^2\, \lf\|\psi_n - \psi \ri\|_{2}\, \lf\|(-i \nabla + \bbv_j)^2 (\psi_n - \psi) \ri\|_{2} .
	}
Since $\psi_n \to \psi$ and $K_j \psi_n \to K_j \psi$ in $L^2(\R^d)$ for $n \to +\infty$, from here we deduce that
	\begin{equation*}
		\forall\, \varepsilon \!>\! 0\;\; \exists\,N_{\varepsilon}\!>\! 0 \quad \mbox{s.t.} \quad (1 - 2\varepsilon\,\|\nabla \xi\|_{\infty})\, \lf\| \xi (-i \nabla + \bbv_j) (\psi_n - \psi)\ri\|_{2} \leqslant \varepsilon^2\,\|\xi\|_{\infty}^2\,,\quad \forall\,n \!>\! N_{\varepsilon}\,,
	\end{equation*}
which implies $\xi (-i \nabla + \bbv_j) (\psi_n - \psi) \to 0$ in $L^2(\R^d,\R^d)$ for all $\xi \in C^{\infty}_{\mathrm{c}}(\R^d \setminus \supp \bbv_j)$. This, in turn, suffices to infer that $(-i \nabla \eta_j) \cdot (-i \nabla + \bbv_j) \psi_n \to (-i \nabla \eta_j) \cdot (-i \nabla + \bbv_j) \psi \in L^2(\R^d,\R^d)$. Summing up, the above arguments entail $K_j(\eta_j \psi_n) \to K_j(\eta_j \psi)$ in $L^2(\R^d)$, ultimately proving that $\eta_j \psi \in \dom(K_j)$. 

Let us finally return to \eqref{eq: Hetajpsin} and notice that, exploiting again basic features of $\bbv_j$ and $\eta_j$, we get $K (\eta_j \psi_n) = K_j(\eta_j \psi_n)$. Since we have just proved that the sequence $K_j(\eta_j \psi_n)$ is convergent in $L^2(\R^d)$ for $n \to +\infty$ and given that $K$ is a closed operator by definition, we readily obtain that $\eta_j \psi \in \dom(K)$ thus concluding the proof.
\end{proof}

	\begin{lemma}\label{lemma:BG24}
		Assume hypotheses \emph{(H1)~--~(H4)} and let $\{\eta_j\}_{j \in J_0}$ be a family of cut-off functions as in \cref{lemma:etaj}. Then, for all $j \in J_0$, the following implications hold true:
		\begin{gather}	
			\psi \in \dom(K_{j}^{*})	\quad \Longrightarrow \quad (1-\eta_j) \psi \in \dom(K_{j})\,; \label{eq:unomenoetajpsi}\\
			\psi \in \dom(K^{*})	\quad \Longrightarrow \quad \lf(1 - {\tx \sum_{j \in J_0}} \eta_j \ri) \psi \in \dom(K)\,. \label{eq:unomenoetapsi}
		\end{gather}
	\end{lemma}
	
\begin{proof}
We prove \eqref{eq:unomenoetajpsi}. The implication in \eqref{eq:unomenoetapsi} can be derived by similar considerations.
Let $\psi \in \dom(K_{j}^{*})$ and set $\xi_j := 1 - \eta_j \in C^{\infty}_{\mathrm{c}}(\R^d \setminus \supp \bbv_j)$. By \eqref{eq:etajpsiDHjst} we readily infer $\xi_j \psi \in \dom(K_{j}^{*})$. Furthermore, noting that $\supp \xi_j \subset \R^d \setminus \supp \bbv_j$, by \eqref{eq:Hjetajpsi} we deduce $K_{j}^{*}(\xi_j \psi) = - \Delta(\xi_j \psi) \in L^2(\R^d)$, which in turn implies $\xi_j \psi \in H^2(\R^d)$ by elliptic regularity.

Next, we proceed to construct a sequence of smooth approximants converging to $\xi_j \psi$ in the graph-norm topology induced by $K_j$. To this purpose, let $\zeta,Z \in C^{\infty}_{\mathrm{c}}(\R^d)$ be such that
	\begin{gather*}
		\zeta \geqslant 0\,, \quad\qquad
		\zeta(\mathbf{x}) = 0, \quad \mbox{ for } |\mathbf{x}| \geqslant 1 \,, \quad\qquad
		\int_{\R^d} \, \diff\mathbf{x}\; \zeta(\mathbf{x}) = 1\,,
		\\
		Z \geqslant 0\,, \qquad 
		Z(\mathbf{x}) = \begin{cases} 1 & \mbox{for }  |\mathbf{x}| \!\leqslant\! 1 \,  \\ 0 & \mbox{for } |\mathbf{x}| \!\geqslant\! 2 \, 
		\end{cases},
		 \qquad 
		\sup_{0 \leqslant |\alpha| \leqslant 2} \|\partial^{\alpha} Z\|_{\infty} \leqslant M, \quad \mbox{for some finite } M > 0 \,,
	\end{gather*}
and set, for $n \in \mathbb{N}$,
	\begin{equation*}
		\zeta_n(\mathbf{x}) := n^d\,\zeta(n \mathbf{x})\,, \quad\qquad
		Z_n(\mathbf{x}) := Z(\mathbf{x}/n)\,, \quad\qquad
		f_{j,n} := \zeta_n \ast \lf( Z_n \xi_j \psi \ri) .
	\end{equation*}
It can be checked that $f_{j,n} \in C^{\infty}_{\mathrm{c}}(\R^d \setminus \supp \bbv_j)$, for any $n$ large enough, and that $f_{j,n} \to \xi_j \psi$ in $H^2(\R^d)$ as $n \to +\infty$.
In particular, we have $\xi_j \psi \in H^2_0(\R^d \setminus \supp \bbv_j)$ and
	\begin{equation*}
		K_j f_{j,n} = - \Delta f_{j,n} \, \xrightarrow[n \to +\infty]{L^2(\R^d)} \, -\Delta (\xi_j \psi)\,.
	\end{equation*}
Since $K_j$ is closed, the above arguments suffice to infer that $\xi_j \psi \in \dom(K_j)$ and $K_j (\xi_j \psi) = - \Delta(\xi_j \psi)$.
\end{proof}

We are now in the position to prove our main result.

\begin{proof}[Proof of \cref{thm:defind}]
We systematically refer to the families of cut-off functions $\{\rho_j\}_{j \in J_0}$, $\{\eta_j\}_{j \in J_0}$ and $\{\tilde{\eta}_j\}_{j \in J_0}$ introduced in \cref{lemma:etaj}. In view of \cref{lemma:SKatosmall}, we henceforth consider the operators $K$ and $K_j$, $j \in J_0$, in place of $H$ and $H_j$, respectively, and proceed to employ the auxiliary \cref{lemma:BG22,lemma:BG23,lemma:BG24} without further concern. Recalling the basic identities reported in \eqref{eq:npnm}, we introduce the shorthand notations
	\begin{equation*}
		n = n_{+}(K)\,, \qquad\quad n_j = n_{+}(K_j)\,, \quad \mbox{for $j \in J_0$}\,.
	\end{equation*}
We remark that, for any given $n \in \mathbb{N} \cup \{\infty\}$, there certainly exists a set $\big\{\Phi_{\ell}\big\}_{1 \leqslant \ell \leqslant 2 n}$ of elements in $\dom(K^{*})$ which are linearly independent modulo $\dom(K)$, namely,
	\begin{equation}\label{eq:linindH}
		{\tx \sum_{\ell \,=\, 1}^{2 n}}\, b_{\ell}\, \Phi_{\ell} \in \dom(K), \quad \mbox{for some } \lf\{ b_{\ell} \ri\}_{\ell \in \{1, \ldots, 2n\}} \subset \mathbb{C} 
		\quad \Longleftrightarrow \quad 
		b_{\ell} = 0, \quad   \forall 1 \leqslant \ell \leqslant 2 n \,.
	\end{equation}
Similarly, for any fixed $j \in J_0$ and for any given $n_j \in \mathbb{N} \cup \{\infty\}$, there exists a set $\big\{\Psi_{j,\ell_j}\big\}_{1 \leqslant \ell_j \leqslant 2 n_j}$ of elements in $\dom(K_{j}^{*})$ which are linearly independent modulo $\dom(K_j)$, namely,
	\begin{equation}\label{eq:linindHj}
		{\tx \sum_{\ell_j \,=\, 1}^{2 n_j}}\, b_{j,\ell_j} \Psi_{j,\ell_j} \in \dom(K_j), \quad \mbox{for some } \lf\{ b_{j,\ell_j} \ri\}_{\ell_j \in \{1, \ldots, 2n\}} \subset \mathbb{C} 
		\quad \Longleftrightarrow \quad 
		b_{j,\ell_j} \!= 0, \quad	\forall 1 \leqslant \ell_j \leqslant 2 n_j \,.
	\end{equation}
We proceed to discuss in separate steps two complementary inequalities, ultimately implying 
	\begin{equation}\label{eq:npmSumK}
		n = {\tx \sum_{j \in J_0}}\, n_j\,,
	\end{equation}
whence the thesis \eqref{eq:npmSum}.

\noindent
{\sl i) Proving that $n \leqslant {\tx \sum_{j \in J_0}} n_{j}$.}
Let $\big\{\Phi_{\ell}\big\}_{1 \leqslant \ell \leqslant 2 n} \!\subset\! \dom(K^{*})$ be any family fulfilling \eqref{eq:linindH}. \cref{lemma:BG22} implies
	\begin{equation*}
		\eta_j \Phi_{\ell} \in \dom(K_{j}^{*}) \cap \dom(K^{*}), \qquad \mbox{for all } j \!\in\! J_0 \mbox{ and } 1 \leqslant \ell \leqslant 2 n\,.
	\end{equation*}
Assume that 	
	\begin{equation}\label{eq:linindetaphi}
		{\tx \sum_{\ell = 1}^{2n}}\, b_{\ell}\, \eta_j \Phi_{\ell} \in \dom(K_j), \quad \mbox{for some } \lf\{ b_{\ell} \ri\}_{\ell \in \{1, \ldots, 2n\}} \subset \mathbb{C}\,.
	\end{equation}
This requirement and \cref{lemma:BG23} (see also \cref{rem:tildeeta}) entail ${\tx \sum_{\ell = 1}^{2n}}\, b_{\ell}\, \eta_j \Phi_{\ell} = \tilde{\eta}_j \lf({\tx \sum_{\ell = 1}^{2n}}\, b_{\ell}\, \eta_j \Phi_{\ell}\ri) \!\in \dom(K)$ and, given that the supports of the cut-off functions $\eta_j$ are disjoint,
	\begin{equation}\label{eq:etaphij1}
		{\tx \sum_{j \in J_0} \sum_{\ell = 1}^{2n}}\, b_{\ell}\, \eta_j \Phi_{\ell} \in \dom(K)\,.
	\end{equation}
On the other hand, by \cref{lemma:BG24}, we get
	\begin{equation}\label{eq:etaphij2}
		\lf( 1 - {\tx \sum_{j \in J_0}} \eta_j\ri) \Phi_{\ell} \in \dom(K)\,.
	\end{equation}
Summing up, \eqref{eq:etaphij1} and \eqref{eq:etaphij2} give
	\begin{equation*}
		{\tx \sum_{\ell = 1}^{2n}}\, b_{\ell}\, \eta_j \Phi_{\ell} 
		= {\tx \sum_{j \in J_0} \sum_{\ell = 1}^{2n}}\, b_{\ell}\, \eta_j \Phi_{\ell} + {\tx \sum_{\ell = 1}^{2n}}\, b_{\ell} \lf( 1 - {\tx \sum_{j \in J_0}} \eta_j\ri) \Phi_{\ell} \in \dom(K)\,.
	\end{equation*}
Due to \eqref{eq:linindH}, the above condition can be fulfilled only if $b_{\ell} = 0$ for all $1 \leqslant \ell \leqslant 2n$. This means that $\{\eta_j \Phi_{\ell} \}_{1 \leqslant \ell \leqslant 2n} \subset \dom(K_j^{*})$ are linearly independent modulo $\dom(K_j)$. As a consequence, there exist $\lf\{ b_{j,\ell_j} \ri\}_{\ell_j \in \{1, \ldots, 2n\}}$ not identically zero and $\varphi_j \in \dom(K_j)$ such that
	\begin{equation*}
		\eta_j \Phi_{\ell} = {\tx \sum_{\ell_j = 1}^{2 n_j}}\, b_{j,\ell_j} \Psi_{j,\ell_j} + \varphi_j \,.
	\end{equation*}
The above arguments ultimately prove that
	\begin{equation*}
		2n \leqslant 2\, {\tx \sum_{j \in J_0}} \dim\big[\dom(K_j^{*})/\dom(K_j)\big] = 2\, {\tx \sum_{j \in J_0}} n_j\,.
	\end{equation*}

\noindent
{\sl ii) Proving that $n \!\geqslant\! {\tx \sum_{j \in J_0}} n_{j}$.}
For any fixed $j \!\in\! J_0$, let $\big\{\Psi_{j,\ell_j}\big\}_{1 \leqslant \ell_j \leqslant 2 n_j} \!\subset\! \dom(K_{j}^{*})$ be some given set fulfilling the condition \eqref{eq:linindHj}. By \cref{lemma:BG22}, we readily infer that
	\begin{equation*}
		\eta_j \Psi_{j,\ell_j} \!\in \dom(K_{j}^{*}) \cap \dom(K^{*}), \qquad
		\mbox{for all } j \in J_0 \mbox{ and } 1 \leqslant \ell_j \leqslant 2 n_j \,.
	\end{equation*}
Assume now that
	\begin{equation}\label{eq:linindetapsi}
		{\tx \sum_{j \in J_0}\, \sum_{\ell_j = 1}^{2 n_j}}\, b_{j,\ell_j} \eta_j \Psi_{j,\ell_j} \in \dom(K), \quad \mbox{for some } \lf\{ b_{j,\ell_j} \ri\}_{\ell_j \in \{1, \ldots, 2n\}} \subset \mathbb{C} \,.
	\end{equation}
Since the supports of the cut-off functions $\{\eta_j\}_{j \in J_0}$ are disjoint, this condition entails
	\begin{equation*}
		\eta_j \,{\tx \sum_{\ell_j = 1}^{2 n_j}}\, b_{j,\ell_j} \Psi_{j,\ell_j} \in \dom(K)\,, \qquad \mbox{for all $j \in J_0$}\,,
	\end{equation*}
which, by \cref{lemma:BG23}, implies in turn
	\begin{equation}\label{eq:etapsij1}
		\eta_j \,{\tx \sum_{\ell_j = 1}^{2 n_j}}\, b_{j,\ell_j} \Psi_{j,\ell_j} 
		= \tilde{\eta}_j \lf( \eta_j \,{\tx \sum_{\ell_j = 1}^{2 n_j}}\, b_{j,\ell_j} \Psi_{j,\ell_j} \ri) \in \dom(K_j)\,.
	\end{equation}
On the other hand, \cref{lemma:BG24} yields
	\begin{equation}\label{eq:etapsij2}
		(1-\eta_j) \,{\tx \sum_{\ell_j = 1}^{2 n_j}}\, b_{j,\ell_j} \Psi_{j,\ell_j} \in \dom(K_j)\,.
	\end{equation}
Summing up, \eqref{eq:etapsij1} and \eqref{eq:etapsij2} give
	\begin{equation}
		{\tx \sum_{\ell_j = 1}^{2 n_j}}\, b_{j,\ell_j} \Psi_{j,\ell_j} \in \dom(K_j)\,,
	\end{equation}
which, on account of condition \eqref{eq:linindHj}, can be fulfilled if and only if $b_{j,\ell_j} \!= 0$ for all $1 \leqslant \ell_j \leqslant 2 n_j$. In view of \eqref{eq:linindetapsi}, this shows that $\big\{\eta_j \Psi_{j,\ell_j}\big\}_{j \in J_0, 1 \leqslant \ell_j \leqslant 2 n_j}$ is a set of elements in $\dom(K^{*})$ which are linearly independent modulo $\dom(K)$. In particular, it follows that
	\begin{equation*}
		2n = \dim\big[\dom(K^{*})/\dom(K)\big] \geqslant 2\, {\tx \sum_{j \in J_0}} n_j\,.
	\end{equation*}	
\end{proof}

\end{document}